\documentclass[aps,prl,twocolumn,groupedaddress]{revtex4}
\usepackage{latexsym}
\usepackage[xdvi]{graphicx}
\usepackage{amssymb,amsmath}

\begin{document}

\title{Magnetic properties of Mn-doped Bi$_2$Se$_3$ compound: \\
temperature dependence and pressure effects}

\author{A.S.~Panfilov}
\email{panfilov@ilt.kharkov.ua}
\affiliation{B. Verkin Institute for Low Temperature Physics and
Engineering, National Academy of Sciences of Ukraine, 61103 Kharkov, Ukraine}

\author{G.E.~Grechnev}
\affiliation{B. Verkin Institute for Low Temperature Physics and
Engineering, National Academy of Sciences of Ukraine, 61103 Kharkov, Ukraine}

\author{A.V. Fedorchenko}
\affiliation{B. Verkin Institute for Low Temperature Physics and
Engineering, National Academy of Sciences of Ukraine, 61103 Kharkov, Ukraine}

\author{K Conder}
\affiliation{Laboratory for Developments and Methods, Paul Scherrer Institute, CH-5232 Villigen PSI, Switzerland}

\author{E V Pomjakushina}
\affiliation{Laboratory for Developments and Methods, Paul Scherrer Institute, CH-5232 Villigen PSI, Switzerland}

\begin{abstract}
Magnetic susceptibility $\chi$ of Bi$_{2-x}$Mn$_{x}$Se$_3$
($x = 0.01-0.2$) was measured in the temperature range $4.2-300$ K.
For all the samples, a Curie-Weiss behaviour of
$\chi(T)$ was revealed with effective magnetic moments of Mn ions
corresponding to the spin value S=5/2, which couple
antiferromagnetically with the paramagnetic Curie temperature $\Theta\sim -50$ K.
In addition, for the samples of nominal composition $x$ = 0.1 and 0.2 the effect of
a hydrostatic pressure $P$ up to 2 kbar on $\chi$ has been measured at fixed temperatures 78
and 300 K that allowed to estimate the pressure derivative of
$\Theta$ to be d$\Theta$/d$P\sim-0.8$ K/kbar.
Based on the observed behaviour of $\Theta$ with varied Mn concentration and pressure,
a possible mechanism of interaction between localized Mn moments is discussed.
\end{abstract}


\maketitle

\section{Introduction}

New class of tetradymite semiconductors Bi$_2$Te$_3$, Bi$_2$Se$_3$ and Sb$_2$Te$_3$
is of great attraction due to a variety of unconventional transport properties related
to an anomalous band structure that supports topologically gapless surface states of
the Dirac cone type (see, e.g., Refs. \cite{Xia,Hasan,Qi2} and references therein).
Of particular interest are the effects of magnetic impurities and ferromagnetism
on the surface states \cite{Chen}.
The magnetic dopants such as Fe or Mn are expected to open the gap at the Dirac point
\cite{Chen,Hsieh09,Wray} that would provide a doping control of the many topological
phenomena and could lead to their unique practical applications
in new spintronic or magnetoelectric devices.
In this context, it is of great importance a detailed study of the magnetic properties
of these compounds to shed more light on the nature of magnetic interaction between
dopant magnetic moments and the interplay between their surface and bulk magnetism.

Focusing here on the bulk properties, we would like to mention a number of experimental results
on bulk magnetism which has been reported for magnetically doped topological insulators:
Bi$_{2-x}$M$_x$Te$_3$
(M = Fe \cite{Kulbachinskii01,Kulbachinskii03}, Mn \cite{Hor10,Niu11,Choi04,Choi05}),
Bi$_{2-x}$M$_x$Se$_3$ (M = Fe \cite{Salman,Choi11}, Mn \cite{Choi05,Janicek08,Choi06}, Cr \cite{Choi11}),
Sb$_{2-x}$M$_x$Te$_3$ (M = Fe \cite{Zhou06}, Mn \cite{Choi05,Dyck03,Horak75}, Cr \cite{Dyck05},
V \cite{Dyck02,Dyck02b}).
Another direction of research concerns experimental studies of peculiar magnetic properties
of Mn-doped Bi$_2$Se$_3$ thin films, which were carried out in the recent years
(see detailed papers \cite{Xu12,Bardeleben,Collins} and references therein).
The obtained results revealed a wide variety of magnetic properties depending on the type
of the 3d-dopant as well as the host compound.
For example, the Bi$_{2-x}$Fe$_x$Te$_3$ compounds demonstrate a ferromagnetic ordering
with low-spin state of Fe ions \cite{Kulbachinskii01, Kulbachinskii03}
while in Sb$_{2-x}$Fe$_x$Te$_3$ compounds Fe ions are in a high-spin state
and interact antiferromagnetically \cite{Zhou06}.
Very different behaviour of the magnetic dopant in topological isolators was also confirmed
by  theoretical calculations for some considered systems \cite{Hsieh09,Larson08,Yu10}.
Despite the substantial amount of available information on the bulk magnetism of these compounds,
the inconsistency of some experimental data on the magnetic state of 3d-dopants
\cite{Salman,Choi11} and character of their interaction \cite{Choi04,Dyck03} should be emphasized.
Furthermore, there is no unified viewpoint on the mechanism of interaction between
the magnetic impurities, and the proposed theoretical models of the exchange interaction
\cite{Yu10,Lasia12,Jungwirth02,Zhang12,Rozenberg12} need to be verified by experiment.

In this paper we have attempted to contribute to the problem of diluted magnetic
semiconductors by studying the bulk magnetic properties of Bi$_{2-x}$Mn$_x$Se$_3$ system
in a wide concentration range $(0\le x\le 0.2)$.
The aim of the study was to clarify the magnetic parameters of the system,
as well as to reveal specific features in their concentration dependence
at the limits of solubility of manganese.
In addition, the effects of high pressure on the magnetic susceptibility of the system
were measured here for the first time.
These measurements were motivated by the huge pressure effects on  the transport properties
observed in the pure Bi$_2$Se$_3$ compound \cite{Hamlin12,Kong13}.
It should be pointed out that along with the widely used doping methods,
pressure is a particularly powerful tool for tuning the electronic properties of solids
without introducing a disorder inherent to the chemical substitution.

\section{Experimental}

Single crystals of Bi$_{2-x}$Mn$_x$Se$_3$ with nominal compositions of
$x$ = 0.01, 0.025, 0.05, 0.10 and 0.20 were grown from a melt using the Bridgman method.
Corresponding amounts of Bi, Mn, Se of minimum purity 99.99\%
were mixed and sealed in evacuated silica ampoules.
The ampoules were annealed at 850$^0$C for 16 h for homogenization.
The melt was then cooled down to 550$^0$C at a rate of 4$^0$C/h
and then down to room temperature with a rate of 100$^0$C/h.
Well-formed silvery crystal rods were obtained, which could be easily
cleaved into plates with flat shiny surfaces.
Phase purity of the obtained samples was characterized at room temperature
by powder X-ray diffraction (XRD) using a D8 Advance Bruker AXS
diffractometer with Cu K$_{\alpha}$ radiation.
XRD experiments revealed that the samples consist mainly of the rhombohedral
Bi$_{2-x}$Mn$_x$Se$_3$ phase (space group R-3m) and some amount of impurity phases
MnSe$_2$ and MnSe, which are proportional to the nominal Mn-content.

The temperature dependences of magnetic susceptibility $\chi(T)$ of Mn-doped
Bi$_2$Se$_3$ were measured by a SQUID magnetometer in the temperature range
of $4.2-300$ K at magnetic field of $H=0.05$ T.
A detectable difference between results of the zero field cooled (ZFC)
and field cooled (FC) measurements was not observed.
In addition, the anisotropy of magnetic susceptibility appeared to be small.
As to the manifestation of the impurity phases MnSe$_2$ and MnSe, their relative
contribution to the magnetic susceptibility was assumed to be negligible
because there were no observed evident  peculiarities in the measured $\chi(T)$
at the magnetic transition points intrinsic to these compounds \cite{Peng}.

As seen in figure \ref{X(T)}, for temperatures above 20--50 K the experimental
$\chi(T)$ dependence for all the samples obeys a modified Curie-Weiss law
\begin{equation}
\chi(T)=\chi_0+{C\over{T-\Theta}} \label{X(T)}
\end{equation}
where  $\chi_0$ is temperature independent contribution, $C$ the Curie constant
and $\Theta$ the paramagnetic Curie  temperature.
The corresponding values of the Curie-Weiss law parameters are collected in table \ref{CW}.

\begin{table}[] 
\caption{Curie constant $C$ (in units $10^{-4}$ K$\cdot$emu/g), paramagnetic Curie temperature
$\Theta$ (K) and $\chi_0$ ($10^{-6}$ emu/g) in Bi$_{2-x}$Mn$_x$Se$_3$ samples
at different nominal Mn content $x_{\rm nom}$ and its corrected values $x_{\rm corr}$
(see section {\em Discussion} for details).}
\label{CW}
\begin{center}
\begin{tabular}{llccc}
\hline
$x_{nom}$ & $x_{corr}$ & $C$ & $\Theta$ & $\chi_0$ \\
\hline
0.01             & 0.01  & 0.716  & $-55.5\pm5$  & $-0.32$\\
0.025            & 0.017 & 1.10  & $-57.9\pm5$    & $-0.24$\\
0.05             & 0.05  & 3.25  & $-53.4\pm3$  & $-0.30$\\
0.10~(\#1) & 0.067  & 4.76  & $-57.8\pm3$  & $-0.30$\\
0.10~(\#2) & 0.105 & 7.25  & $-54.9\pm3$  & $-0.30$\\
0.10~(\#3) & 0.140 & 9.9  & $-52.9\pm3$  & $-0.35$\\
0.20             & 0.20  & 13.9  & $-54.3\pm3$  & $-0.35$\\
\hline
\end{tabular}
\end{center}
\end{table}

\begin{figure}[]
\begin{center}
\includegraphics[width=0.4 \textwidth]{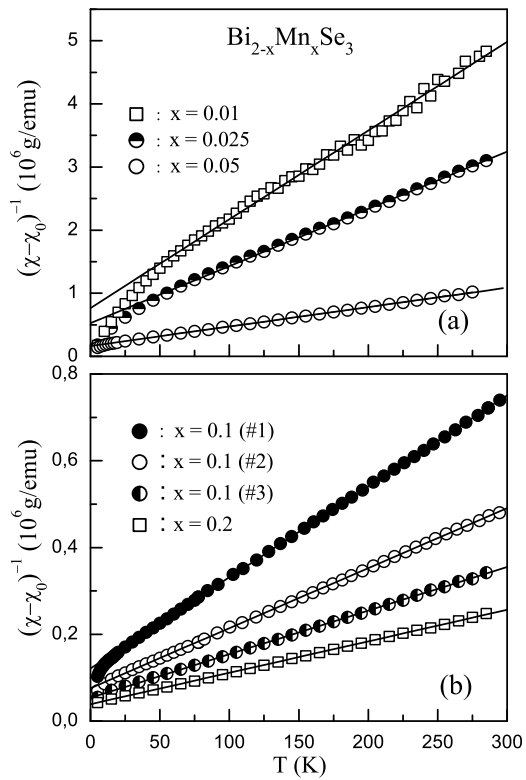}
\caption{Temperature dependence of the
reciprocal magnetic susceptibility $(\chi-\chi_0)^{-1}$ for Bi$_{2-x}$Mn$_x$Se$_3$. The
Curie-Weiss fits are indicated by the solid straight lines.} \label{X(T)}
\end{center}
\end{figure}
\begin{figure}[]
\begin{center}
\includegraphics[width=0.42\textwidth]{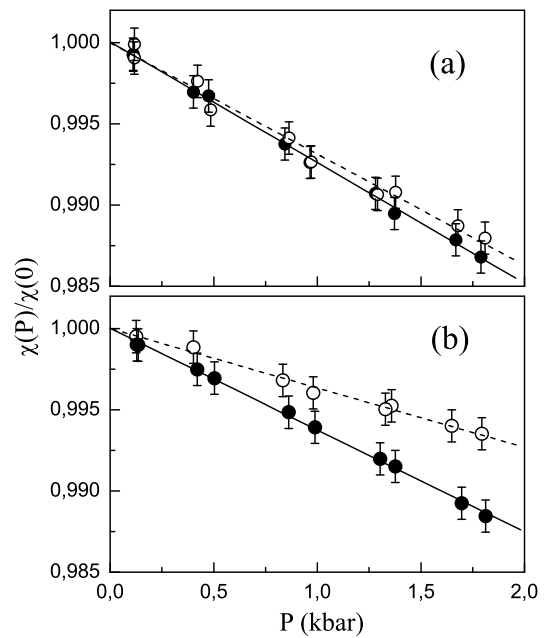}
\caption{Pressure dependencies of the magnetic susceptibility of
Bi$_{1.9}$Mn$_{0.1}$Se$_3$ (\#1) (a) and Bi$_{1.8}$Mn$_{0.2}$Se$_3$ (b) at $T=78$ K (full circles) and 300 K (open circles) normalized to its value at zero pressure.}
\label{X(P)}
\end{center}
\end{figure}

For the samples of Bi$_{2-x}$Mn$_x$Se$_3$ with nominal Mn content
$x=0.1 (\#1)$ and $x=0.20$, the studies of magnetic susceptibility were carried out
under helium gas pressure $P$ up to 2 kbar at fixed temperatures, $T$=78 and 300 K,
using a pendulum-type magnetometer placed directly in the nonmagnetic pressure cell.
The magnetometer construction is similar to that used in Ref.~\cite{Bozorth56}.
The measured sample with typical sizes of about $2.5\times4\times6$ mm$^3$ was placed inside a small
``compensating'' coil located at the lower end of the pendulum rod having a length of about 150 mm.
The working volume is situated between a cone type pole pieces of an electromagnet
in an inhomogeneous magnetic field up to 2 T.
The actual diameter of the poles ends and the gap between them were 100 mm and 40 mm, respectively.
When the magnetic field is applied, the value of a current through the compensating coil,
at which the pendulum comes to its zero position, appears to be a measure of the sample magnetic moment.
In order to measure the pressure effects, the pendulum magnetometer is inserted in the
cylindrical nonmagnetic pressure chamber with outer and inner diameters of 24 mm and 6 mm,
respectively, which is placed in a cryostat.
Measurements were performed at a fixed temperature to eliminate the effects on susceptibility
of temperature changes upon application or removal of pressure.
A detailed description of the device and analysis of the sources of experimental errors
will be published elsewhere.

In our case, the relative errors of measurements of $\chi$ under pressure did not exceed $0.1\%$
for the employed magnetic fields $H\simeq1.7$ T.
The experimental $\chi(P)$ dependencies, normalized to the values of $\chi$ at zero pressure,
are presented in figure \ref{X(P)},
yielding values of the pressure derivative d\,ln$\chi$/d$P$ at different temperatures,
which are listed in table \ref{exp}.

\begin{table}[] 
\caption{Magnetic susceptibility $\chi$ (in units $10^{-6}$ emu/g) and its pressure derivative
d\,ln$\chi$/d$P$ (Mbar$^{-1}$) for Bi$_{2-x}$Mn$_{x}$Se$_3$ samples at different temperatures.}
\vspace{5pt} \label{exp}
\begin{center}
\begin{tabular}{lccc}
\hline
Sample & T (K)& $\chi$ & d\,ln$\chi$/d$P$\\
\hline
$x=0.1 (\#1)$ & 78 & 3.22  & $-7.45\pm0.5$ \\
\vspace {5 pt}    &300 & 1.04  & $-6.85\pm0.5$ \\

$x=0.2$           & 78 & 10.2 & $-6.25\pm0.5$\\
                  &300 & 3.58 & $-3.5\pm0.5$ \\
\hline
\end{tabular}
\end{center}
\end{table}

\section{Discussion}

\begin{figure}[] 
\begin{center}
\includegraphics[width=0.39\textwidth]{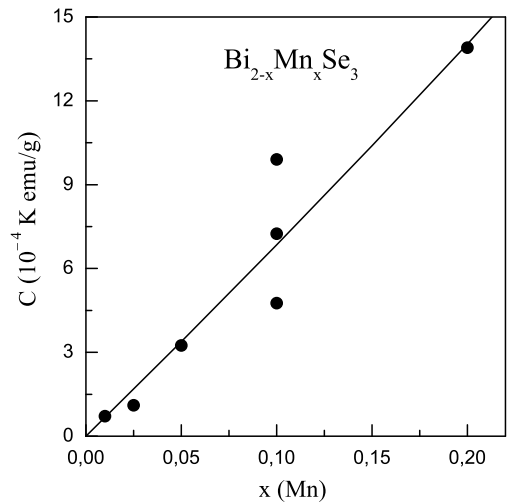}
\caption{Curie constant $C$ for Bi$_{2-x}$Mn$_x$Se$_3$ samples
as a function of Mn content.
Solid line corresponds to description of $C(x)$ assuming effective magnetic moment
value of Mn ion to be $\mu_{\rm eff}=5.92~\mu_{\rm B}$ (S=5/2).} \label{C(x)}
\end{center}
\end{figure}

\begin{figure}[]
\begin{center}
\includegraphics[width=0.4\textwidth]{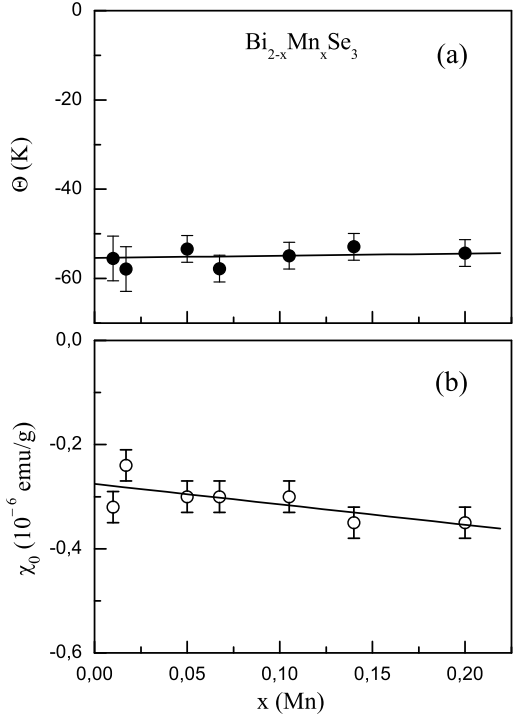}
\caption{Values of paramagnetic Curie temperature $\Theta$ (a) and parameter
$\chi_0$ (b) in Bi$_{2-x}$Mn$_x$Se$_3$ compounds for different Mn content.}
\label{Theta(x)}
\end{center}
\end{figure}

As seen in figure \ref{C(x)}, the concentration dependence of the Curie constant $C(x)$
is close to the linear one determined by the value of effective magnetic moment per Mn atom,
$\mu_{\rm eff}\simeq 5.9~ \mu_{\rm B}$.
This moment corresponds to the spin state of Mn ion S=5/2.
Some deviation of individual data points (e.g., for $x=0.1$) are assumed to be
due to a difference of the actual content of manganese in the samples,
$x_{\rm corr}$, from its nominal value, $x_{\rm nom}$.
Resulted from this assumption corrected values of $x_{\rm corr}$ are given
in table \ref{CW} and used to represent concentration dependencies
of the Curie-Weiss parameters $\Theta$ and $\chi_0$ in figure \ref{Theta(x)}.

As seen in figure \ref{Theta(x)}, the paramagnetic Curie temperature $\Theta$ in
Bi$_{2-x}$Mn$_{x}$Se$_3$ samples shows negative value, which is weakly dependent on
composition and points to a strong antiferromagnetic coupling between magnetic moments of Mn ion.
The averaged over the entire range of concentrations value $\Theta\simeq-55$ K
agrees reasonably with the value of $\Theta\sim-70$ K, which follows from the data
of Ref. \cite{Choi06} for Bi$_{1.97}$Mn$_{0.03}$Se$_3$ sample,
but differs substantially in magnitude from the value $\Theta\sim -0.5$ K reported in
Ref. \cite{Janicek08} for Bi$_{2-x}$Mn$_x$Se$_3$ compounds with $x=0.01$ and 0.02.
It should be noted, however, that the latter estimate was based on an analysis of the
experimental data on magnetic properties at rather low temperatures, $2\le T\le 20$ K.
At these temperatures the result of Ref. \cite{Janicek08} could be influenced by possible manifestations
of the Kondo and crystal electric field effects in magnetism, as well as the contribution of
the impurity ions of manganese.
We believe that this can be a probable explanation for the difference between the estimate
of $\Theta$ in \cite{Janicek08} and our results, which were obtained from analysis of
$\chi(T)$ dependences at the higher temperature range (see figure \ref{X(T)}(a)).
As to the parameter $\chi_0$, its values for Bi$_{2-x}$Mn$_{x}$Se$_3$ compounds
correlate with the reported in literature data on magnetic susceptibility
of pure Bi$_2$Se$_3$, $\chi \simeq -0.3\times10^{-6}$ emu/g \cite{Kulbachinskii03}
and $-0.41\times10^{-6}$ emu/g \cite{Matyas58}.
Thus, this parameter can be identified as the intrinsic susceptibility of Bi$_2$Se$_3$.

\begin{figure}[]
\begin{center}
\includegraphics[width=0.4\textwidth]{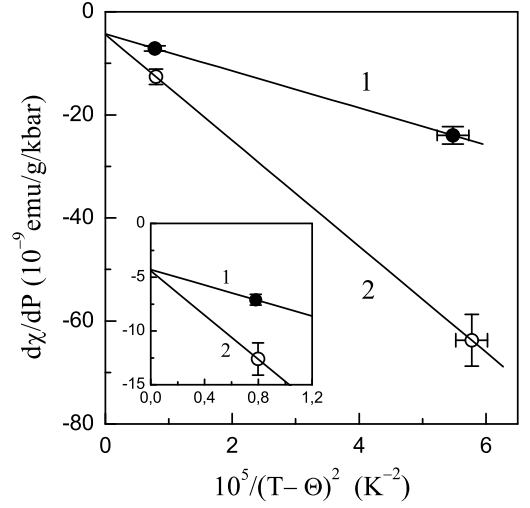}
\caption{Dependence of the pressure derivative d$\chi$/d$P$ for
Bi$_{1.9}$Mn$_{0.1}$Se$_3$ \#1 (1) and Bi$_{1.8}$Mn$_{0.2}$Se$_3$ (2) versus $(T-\Theta)^{-2}$.
The inset shows the data near the origin of coordinates on an expanded scale.} \label{dX-dP}
\end{center}
\end{figure}

Let us now turn to discuss the effects of pressure on magnetic susceptibility.
Neglecting the value of $\chi_0$ in the Curie-Weiss behaviour (\ref{X(T)}) of $\chi(T)$
and assuming the Curie constant $C$ to be pressure independent,
the measured derivative d\,ln$\chi(T)$/d$P$ is obviously governed
by the pressure dependence of the paramagnetic Curie temperature $\Theta$:
\begin{equation}
{{\rm d\,ln}\chi(T)\over{\rm d}P}\approx
{1\over(T-\Theta)}{{\rm d}\Theta\over{\rm d}P}.
\label{dlnX/dP}
\end{equation}
Then substitution in (\ref{dlnX/dP}) the experimental data at $T=78$ K from table \ref{exp}
for the sample with $x=0.2$ yields the value
\begin{equation}
{{\rm d}\Theta\over{\rm
d}P}=-0.8\pm0.1~{\rm K/kbar}\label{dTheta/dP}.
\end{equation}
In more rigorous analysis the term $\chi_0$ has to be taken into account
and then the pressure effect value d$\chi$/d$P \equiv \chi$d\,ln\,$\chi$/d$P$ is given by relation
\begin{equation}
{{\rm d}\chi(T)\over{\rm d}P}={{\rm d}\chi_0\over{\rm d}P}+{C\over(T-\Theta)^2}\times
{{\rm d}\Theta\over{\rm d}P},
\label{dX/dP}
\end{equation}
being a linear function of $(T-\Theta)^{-2}$.
The experimental values of d$\chi$/d$P$ for the samples with $x=0.1$ and 0.2
as a function of $(T-\Theta)^{-2}$ are shown in figure \ref{dX-dP}.
From the slopes of line 1 and 2 in figure~\ref{dX-dP} one can obtain the values of pressure
derivative for paramagnetic Curie temperature to be d$\Theta$/d$P=-0.75\pm0.05$ K/kbar and
$-0.74\pm0.05$ K/kbar for Bi$_{1.9}$Mn$_{0.1}$Se$_3$ and Bi$_{1.8}$Mn$_{0.2}$Se$_3$, respectively.
For both compounds, the corresponding pressure derivative of the relative change in $\Theta$ amounts to
\begin{equation}
{\rm d\,ln}\,\Theta/{\rm d}P=14\pm 1~ {\rm Mbar}^{-1}.
\label{dlnTheta/dP}
\end{equation}
The observed substantial pressure effect on $\Theta$, d$\Theta$/d$P=-0.75$ K/kbar,
correlates both in sign and magnitude with the reported in Ref. \cite{Dyck06b} value of
pressure effect on the Curie temperature $T_{\rm C}$ for V-doped Sb$_2$Te$_3$ compound,
d$T_{\rm C}$/d$P=-1.3$ K/kbar, that suggests a common nature of magnetic interactions in both systems.

Another principal parameter resulted from the data represented in figure~\ref{dX-dP}
is the pressure derivative of the value $\chi_0$.
It was estimated by extrapolation  $(T-\Theta)^{-2}\to 0$ to be
\begin{equation}
{\rm d}\chi_0/{\rm d}P=-4.3\pm0.5~~ {\rm and} -4.4\pm1.5
\end{equation}
in units of $10^{-9}$ (emu/g)/kbar for the samples with $x=0.1$ and $0.2$, respectively.
Using this estimate and the value $\chi_0\simeq 0.3\times10^{-6}$ emu/g from table \ref{CW}, one obtains
\begin{equation}
{\rm d\,ln}\,\chi_0/{\rm d}P=14\pm3~ {\rm Mbar}^{-1}.
\label{dlnchi_0}
\end{equation}
As can be seen from Eqs. (\ref{dlnTheta/dP}) and (\ref{dlnchi_0}),
both $\Theta$ and $\chi_0$ parameters show significant pressure effects,
which are  similar in magnitude and correlate with the experimentally observed
changes under pressure of the carrier density $n$
\begin{equation}
{\rm d\,ln}\,n/{\rm d}P=15\div20~{\rm Mbar}^{-1},
\label{dln<n>/dP}
\end{equation}
resulted from the initial slope of the $n(P)$ dependence in Bi$_2$Se$_3$ \cite{Hamlin12,Kong13}.

It is quite curious that the observed Curie temperature $\Theta$ itself
and its pressure derivative do not notably depend on the Mn concentration.
This allows to presume that Mn doping effects have the single-site impurity origin,
e.g., of the Kondo type with $T_{\rm K}\approx-\Theta/4\sim 10-15$ K.
The Kondo temperature $T_{\rm K}$ is given by \cite{Coqblin69,Newson93}:
\begin{equation}
k_{\rm B}T_{\rm K}\simeq D~{\rm exp}[-1/JN(E_{\rm F})],
\label{T_K}
\end{equation}
where $D$ is the effective band width, $J$ the interaction parameter between Mn-moment and
conduction electron spins, $N(E_{\rm F})$ density of electronic states at the Fermi energy $E_{\rm F}$.
Then, keeping in mind the strong pressure dependence of the carrier density (\ref{dln<n>/dP}),
one can assume the dominant contribution in the pressure effect on $T_{\rm K} (\Theta)$
to be due to the increasing under pressure the density of states  $N(E_{\rm F})$,
which is proportional to $n^{1/3}$ for the parabolic band.

An alternative mechanism for the observed effects of manganese doping involves
the formation of the Mn-enriched clusters with similar magnetic properties
instead of a uniform distribution of the manganese atoms throughout the crystal.
For this case, the paramagnetic Curie temperature $\Theta$ in the cluster can be determined
by an indirect exchange interaction between Mn-moments mediated by conduction electrons.
In the framework of the conventional RKKY approach \cite{Mattis65},
the main functional dependence of $\Theta$ may be given by:
\begin{equation}
\Theta\propto J^2n^{4/3}F(n)
\label{Theta_RKKY}
\end{equation}
where $J$ is the exchange coupling parameter between Mn-moments and conduction electron spins,
$F(n)$ the RKKY function.
As seen from Eq. (\ref{Theta_RKKY}), the strong pressure dependence of $\Theta$
could be governed by the pressure effect on the carrier density $n(P)$.

It should be admitted that in order to make a choice between the above proposed approaches,
we need a more rigorous quantitative analysis, which represent a challenging task
and is not the subject of this paper.

Let us conclude by discussing the derived pressure effect on the parameter $\chi_0$,
which can be identified as the intrinsic susceptibility of Bi$_2$Se$_3$, $\chi$.
It can generally be represented as the sum:
\begin{equation}
\chi=\chi_{\rm P}+\chi_{\rm VV}+\chi_{\rm LP}+\chi_{\rm ion}.
\label{X_t}
\end{equation}
Here $\chi_{\rm P}$ is the Pauli spin susceptibility,
$\chi_{\rm VV}$ a generalization of the Van Vleck orbital paramagnetism,
$\chi_{\rm LP}$ the Landau--Peierls diamagnetism and
$\chi_{\rm ion}$ the Langevin diamagnetism of closed ion shells.
For a parabolic band, the Pauli paramagnetism and Landau--Peierls diamagnetism
are described by well known relations \cite{Wilson53}
\begin{equation}
\chi_{\rm P}={4m^*\mu_{\rm B}^2\over h^2}(3\pi^2n)^{1\over3},
~~~~\chi_{\rm LP}=-{4m_0^2\mu_{\rm B}^2\over 3m^*h^2}(3\pi^2n)^{1\over3},
\label{X_c}
\end{equation}
where $m^*$ is the effective mass of band electrons.
Further, using for Bi$_2$Se$_3$ the estimate of $m^*=0.11\div0.18~m_0$
\cite{Hyde74,Kulbachinskii99,Sugama01}, one can conclude that $\chi_{\rm LP}>>\chi_{\rm P}$.
The value of orbital susceptibility was calculated to be
$\chi_{\rm VV}\simeq0.145\times 10^{-6}$ emu/g \cite{Yu10},
and the Langevin term, $\chi_{\rm ion}\simeq -0.11\times 10^{-6}$ emu/g,
results from the data for the ion susceptibilities of Bi$^{5+}$ and Se$^{6+}$ \cite{Selwood56}.
Based on the above estimates and the measured value of magnetic susceptibility for Bi$_2$Se$_3$,
$\chi\simeq 0.30\times 10^{-6}$ emu/g, we have evaluated the Landau-Peierles term to be
\begin{equation}
\chi_{\rm LP}\simeq -0.34\times 10^{-6}~{\rm emu/g}.
\label{X_LP}
\end{equation}
Because the orbital contribution is weakly dependent on pressure and the Pauli paramagnetism
is relatively small, it is reasonable to assume that the Landau-Peierles term gives the main
contribution to the measured pressure effect, i.e. d$\chi_0\simeq{\rm d}\chi_{\rm LP}$ and
\begin{equation}
{{\rm d\,ln}\chi_{\rm LP}\over{\rm d}P}\equiv {1\over\chi_{\rm LP}}
{{\rm d}\chi_0\over{\rm d}P}\simeq 12.5~{\rm Mbar^{-1}}.
\label{dX_LP}
\end{equation}
As it follows from Eq. (\ref{X_c}),
\begin{equation}
{{\rm d\,ln}\chi_{\rm LP}\over{\rm d}P}={1\over 3}{{\rm d\,ln}\,n\over{\rm d}P}-{{\rm d\,ln}\,m^*\over{\rm d}P}.
\label{dlnX_LP}
\end{equation}

\begin{figure}[] 
\begin{center}
\includegraphics[width=0.4\textwidth]{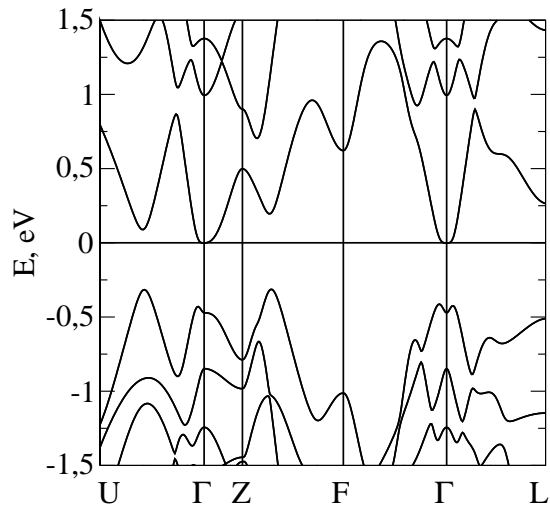}
\caption{Band structure of Bi$_2$Se$_3$ along the symmetry
directions of the rhombohedral Brillouin zone.
The Fermi level is indicated by the horizontal line at $E$=0.} \label{bs2}
\end{center}
\end{figure}

In order to estimate the pressure dependence of $m^*$,
we have carried out the calculations of the volume dependent band structure of Bi$_2$Se$_3$,
which is presented in figure \ref{bs2}.

The calculations were performed by employing a relativistic full-potential LMTO method
(FP-LMTO, RSPt code \cite{wills10,grechnev09,rspt}) within the generalized gradient
approximation \cite{gga} of the density functional theory.
The effective mass of electrons was obtained using a parabolic band model at the conduction
band minimum by a parabolic fitting to the band dispersion $E(k)$ along different directions
around the $\Gamma$ symmetry point of the Brillouin zone (see figure \ref{bs2})
and employing the relation $m^*= 1/({\partial}^2E/{\partial}k^2)$ (in atomic units).
The calculated average value of the effective mass, $m^*\simeq 0.12~m_0$,
is found to increase with volume $V$, and the corresponding volume derivative of $m^*$
appeared to be d\,ln\,$m^*$/d\,ln$V\simeq 3$.
The obtained values are assumed to be valid for small carrier density
inherent to real samples of the Bi$_2$Se$_3$ compound.
Taking into account the bulk modulus value $B\simeq0.5$ Mbar for Bi$_2$Se$_3$ \cite{Zhao13},
we have estimated the pressure derivative of $m^*$ in (15) as d\,ln\,$m^*$/d$P\simeq-6$ Mbar$^{-1}$.
By substituting this estimate to Eq. (\ref{dlnX_LP}),
together with the first term, (1/3)d\,ln\,$n$/d$P=5\div 7$ Mbar$^{-1}$ \cite{Hamlin12,Kong13},
we have obtained a reasonable agreement of the model (\ref{dlnX_LP}) with the experimental result (\ref{dX_LP}).

\section{Summary}

In conclusion, the temperature dependence of magnetic susceptibility for
Bi$_{2-x}$Mn$_x$Se$_3$ compounds was measured for the concentration range up to $x=0.2$.
For all the samples, the experimental data obey a Curie-Weiss law
with the effective magnetic moment of Mn ions corresponding to the spin value S=5/2
and antiferromagnetic coupling of the moments.
For Mn-rich samples, the dependence of magnetic susceptibility
on the pressure was measured for the first time.

The most surprising result is that the observed Curie-Weiss parameters,
$\Theta$ and $\chi_0$, and their available pressure derivatives
do not depend notably on the Mn concentration.
This behaviour suggests that the effects of Mn doping are the single-site impurity origin
of the Kondo type or these effects involve the formation of the Mn-enriched clusters
of the similar magnetic properties instead of a uniform distribution
of the manganese atoms throughout the crystal.
In both approaches, the large pressure effects in magnetism of the system
points to their strong correlation with the literature data on the pressure
dependence of the carrier density in Bi$_2$Se$_3$.

\section*{Acknowledgments}
The authors thank Dr. G.P. Mikitik for fruitful discussions and comments.


\end{document}